# A semi-analytic reconstruction method for Diffuse Optical Tomography

Tapan Das[a,*], P. K. Dutta[a]

*Electrical Engineering Department, Indian Institute of Technology Kharagpur, Kharagpur, 721302, West Bengal, India*

**Abstract**

The Diffuse Optical Tomography (DOT) has received considerable attention in the recent years in the field of biomedical imaging and disease detection. However, imaging through highly diffusive medium is a challenge and stability is always an issue due to the inverse problem. Here a non-linear continuous wave (CW) semi-analytic reconstruction method is discussed that used curved-beam paths for tomographic imaging with no assumption on inclusion, unlike the iterative methods. The non-linear Rosenbrock's function is used to approximate the paths followed by majority photons as curved ones. The modified Beer-Lambert Law (MBLL) in proposed differential form is used to calculate the relative absorption coefficient ($\mu_a$) of the all the available photon paths. The computed values are back-projected along these channels and serve as the basis for image reconstruction without solving the inverse problem. For three-dimensional (3-D) imaging, measurements at three different depths covering the entire depth of the phantom are taken. These slices are stacked together followed by interpolation to form the volumetric image of the phantom to complete the tomographic imaging in its true sense. Numerical simulations, wax phantom experiments with different geometries and contrast are carried out with satisfactory results. This semi-analytic DOT reconstruction method is simple and efficient and is suitable for real-time applications that require fast absorption image reconstruction. The method is compared with the Greedy algorithms for further validation. Also, different performance evaluation matrices are estimated to assess the accuracy of the reconstructed images and the results are rather satisfactory.

## 1. Introduction

The Diffuse Optical Tomography (DOT) is a non-invasive and non-ionizing imaging technique that has got considerable attention in recent years. It uses near-infrared (NIR) light for determination of optical properties of tissue from boundary measurement [1]. The major application areas of DOT are the brain, breast, and finger joint imaging for diseases detection [2-5]. The highly scattering nature of tissue makes the image reconstruction an underdetermined and ill-posed problem that require complex reconstruction algorithms [5,6]. The conventional DOT imaging methods solve the reconstruction problem iteratively, and through different linearization techniques [7].

*Corresponding author.
   Email address:* tapandas@iitkgp.ac.in (Tapan Das)

However, the iterative methods calculate the forward diffusion equation repeatedly that adds to the computational complexity. Different linearization techniques are used in literature to overcome the nonlinearity present in the inverse problem [8,9]. However, it fails when the contrast between the unknown optical properties and homogeneous background exceeds born estimated limit [10]. Thus, the solution is unstable, or it may not even have a unique solution [9]. To obtain unique solution different regularization techniques are also incorporated in the reconstruction schemes as discussed in the literature [11,12]. Compressive sensing (CS) based DOT reconstruction method has received considerable attention in recent years because of its efficiency. However, these DOT algorithms have to solve the inverse problem. Thus, stability is always an issue and due to complex mathematics involved in the reconstruction the computational overhead is quite high [13]. Non-recursive linear model that estimates photon paths using a linear relationship is already reported in the literature [14] and produces fast and usable images without any inverse transforms. Thus, stability is not an issue. However, linear approximations result in distortion in the reconstructed images. The motivation behind this work is for developing a CW low cost non-linear 3-D tomographic imaging method with no assumption on perturbation and computational efficiency suitable for practical DOT applications. Here the focus is on the reconstruction of absorption images only as it is considered to diagnose the affected tissue volumes [15]. The technique is used for 3-D reconstruction of different geometry phantoms with variation in the contrast between the background and the target.

The non-linear Rosenbrock's banana function is fitted to each source- detector channel by curve fitting to approximate the average photon paths from the source to a detector as curved one [16]. Here we reconstructed 3-D volumetric images based on MBLL in differential form without employing any inverse transforms. In general, the difference data are obtained either by using wavelength-difference imaging or state-difference imaging. Here in this proposed method, the difference data is calculated between two-photon channels estimated by fitting of the Rosenbrock's banana function. The photon channel with the least OD value and smallest source-detector separation is considered as the reference channel. This technique eliminates common system noise of the paths corresponding to the detector placed closest to the source, which is not probing the tissue. This normalization acts as a pre-processing step and helps to eliminate reconstruction artifacts near the tissue boundaries. This method considers each photon channel to have different $\mu_a$ while scattering coefficient ($\mu_s'$) is assumed to be constant throughout the imaging volume. Also, the total mean path length (<L>) in the MBLL is replaced by the estimated average path length ($L^i$) for any $i^{th}$ photon channel. This replacement reduces the partial volume effect associated with the MBLL due to the assumption that changes in $\mu_a$ are spatially uniform over the entire measurement volume. Now, the MBLL in this proposed form is applied to the reference channel and all the other available channels to calculate relative absorption coefficient ($\mu_a$) between them and this difference measurement also nullified the effect of scattering loss factor ($G$) in the calculations. Thus, there is no need for any additional baseline measurement to obtain the absorption value in this technique. The method also does not assume inclusion location, shape, and size unlike most of the iterative DOT methods. The calculated $\mu_a$ values are back projected along these curved photon channels between source-detector pairs for the two-dimensional (2-D) image reconstruction without solving the inverse problem. The use of the curved photon paths as a basis for reconstruction helped in correcting the image distortion associated with linear methods [14]. Now for 3-D volumetric imaging, measurements at three different depths covering the entire depth of the phantom are considered. The corresponding 2-D slices are reconstructed using this method and the same are stacked together followed by interpolation to reconstruct the complete volumetric image of the phantom [17]. One of the major advantages of this non-linear technique is the presence of hardly any ghost object in the reconstructed image, unlike iterative methods. Thus, this relatively fast and accurate imaging method is suitable real-time monitoring of absorption images. To show the usability of this method numerical simulation using TOAST++ [18] and experiments with different geometry (rectangular and cylindrical) wax phantoms are carried out. Also, targets with varying values of $\mu_a$ are used to see the contrast in the reconstructed images. The results show good accuracy in locating targets with satisfactory contrast. Another experiment on a wax slab phantom is carried out having inclusion at the center covering the entire length of the phantom. The experiment showed there is considerable amount of light attenuation due to the presence of inclusion in the curved photon paths.

A comparative study is carried out with the Greedy algorithms to assess performance of this reconstruction method. The different Greedy algorithms used in DOT are discussed in details in literature [13]. The advantage of the Greedy algorithm used here is that the forward diffusion equation is not to be solved repeatedly. This curved-beam method is compared with the conventional MMV algorithms like the compressive sampling matching pursuit (CoSaMP) and the subspace augmented-multiple signal classification (SA-MUSIC) [19]. This non-linear, non-recursive method is computationally efficient with good recovery accuracy. And is a straightforward approach for DOT reconstruction compared to Greedy algorithms that involve complex mathematics. The performance evaluation metrics like the structural similarity index (SSIM) [20], mean squared error (MSE), and peak signal-to-noise ratio (PSNR) are evaluated for performance assessment and comparison of reconstructed images. Thus, this method alternative approach for DOT reconstruction is suitable for real-time applications that require fast monitoring of absorption coefficient. However, this non-linear semi-analytic method can be improved further by estimating the curved photon channels as a cloud of photon probability. The future work will be carried out in this direction.

## 2. Methodology

*2.1 Rosenbrock's Banana function and its fitting*

Light propagation in tissue is a complicated process, and undergoes multiple scattering before some part of it is reflected back to the surface, to be detected by detectors placed at the boundary. It is reported in the literature that majority photons take a curved path between the source-detector pairs [21,22]. So, the idea here is to fit a non-linear semi-analytic function as shown in Eq. (1), to each of the source-detector channel data by curve fitting. This curve fitting estimates the photon path followed by the majority of photons from a light source to all the detectors as curved one rather than hypothesizing it to be linear [14]. The Rosenbrock's banana function is given as:

$$Z = f(x,y) = 100(y - x^2)^2 + (a - x)^2 \tag{1}$$

The minimum value of this function $Z = f(x, y)$ is obtained by setting gradient to zero and solving the system of two coupled equations:

$$\frac{df}{dx} = -2(a - x) - 400(y - x^2)x = 0 \tag{2}$$

$$\frac{df}{dy} = 2(y - x^2) = 0 \tag{3}$$

Thus, the minimum point of this function $Z$ is obtained as $(x, y) = (a, a^2)$.

The modified Beer-Lambert law (MBLL) relates incident and detected light intensities in a predominantly scattering medium to optical attenuation for the whole medium as follows

$$A = ln\left(\frac{I_s}{I_d}\right) = \mu_a \times DPF \times d + G = \mu_a <L> + G \tag{4}$$

where, $A$ is the attenuation (defined here as log base e), $I_s$ the incident light intensity, $I_d$ the detected light intensity, $d$ the source-detector distance, $DPF$ is the differential path length factor that accounts for extra distance light has to travel due to scattering between a source-detector pair, $<L>$ is the total mean path length of detected photons, $\mu_a$ is the absorption coefficient of the tissue and $G$ is an unknown scattering loss factor.

In CW measurements $G$ is considered to be known constant, and $\mu_a$ is the unknown parameter to be estimated. Here in this proposed method each source-detector channel is modeled as a local region having different unknown absorption coefficients to be estimated. Also, here, $L^i$ is determined for each channel, and the same replaces the $<L>$ in the conventional MBLL in Eq. (4) before applying it to all the available photon channels. So, MBLL in proposed form is applied to each of these photon channels, and differential measurement is used to calculate the relative $\mu_a$. This is unlike wavelength-difference imaging or state-difference imaging methods. In wavelength-difference imaging data is acquired at two different wavelengths and in state-difference imaging data is collected on the same object before and after some change has occurred, or is induced, in the object's optical properties. Thus, estimation of photon paths has enabled the use of MBLL in its proposed form. And has another advantage of reducing the partial volume effect associated with MBLL due to the assumption that changes in $\mu_a$ are spatially uniform over the entire measurement volume.

Here, the focus is on the reconstruction of 3-D absorption profiles as such value of $L^i$ is approximated as a function of source-detector distance. For CW measurements DPF is not known. The computational model for the homogeneous semi-infinite medium is based on a reasonable approximation of radiative transport equation (RTE) for thick medium called Diffusion equation (DE). The analytical solution of DE is obtained by using a general method of Green's function. The approximate value of DPF or $L^i$ is related to tissue $\mu_a$, $\mu_s'$, and the geometry under consideration as follows [23-25].

$$DPF = \frac{1}{2}\sqrt{\frac{3\mu_s'}{\mu_a}}\left[1 - \frac{1}{\left(1+d\sqrt{3\mu_a\mu_s'}\right)}\right] \tag{5}$$

$$L^i = d\left\{\frac{1}{2}\sqrt{\frac{3\mu_s'}{\mu_a}}\left[1 - \frac{1}{\left(1+d\sqrt{3\mu_a\mu_s'}\right)}\right]\right\} \tag{6}$$

So, with the known values of $\mu_a$, $\mu_s'$, and $d$, the values of *DPF* or $L^i$ are calculated from the above Eq. (5) and (6) that agreed well with the earlier reported values [23]. Our main focus is on the reconstruction of 3-D absorption profiles as such approximate value of DPF is considered. However, advanced methods to measure DPF continuously have also been reported already [26]. In this proposed method the photon paths are estimated as curved one for all the channels by curve fitting of Rosenbrock's function as mentioned above. So, with all the available curved photon channels, the MBLL in the proposed form can be applied to calculate the $A_i$ for an $i^{th}$ channel, using and the already estimated value of $L^i$ for that channel as given by Eq. (6). Here in this proposed form, the $\mu_a$ is assumed to vary from one channel to the other. So, for any two-photon channels $i$ and $j$ the unknown absorption coefficients are considered to be $\mu_a^i$ and $\mu_a^j$ respectively. Thus, $A_i$ and $A_j$ are calculated for both the two channels. Now the attenuation change is calculated from the detected intensity values of the reference channel ($i^{th}$) and all the available channels as follows.

$$\left|A^j - A^i\right| = \ln\left(\frac{I_{d_i}}{I_{d_j}}\right) = L^j\mu_a^j - L^i\mu_a^i \tag{7}$$

where $L^i$, $L^j$ = average path lengths traveled by detected photons inside the respective channels, and *i=1* and *j=2 to 13 or 16* depending on the number of detectors used for different phantom geometries.

Now for the reference channel $L^i$ is very small compared to $L^j$ and $\mu_a^i$ is same as background absorption coefficient ($\mu_o$) because it corresponds to the shortest path and least OD value. Hence in Eq. (7) the second term is neglected to get the approximated value of $\mu_a^j$ of the $j^{th}$ photon channel as shown below.

$$\mu_a^j = \frac{1}{L^j}\ln\left(\frac{I_{d_i}}{I_{d_j}}\right) \tag{8}$$

Thus, from the relative change in *A* obtained from the detected intensity values it is possible to estimate the $\mu_a$ values of the photon channels. The effect of the unknown factor, *G* in the computed $\mu_a$ is also nullified due to the use of MBLL in differential form. Now the optical density (*OD*) values for all the available channels are calculated, and source-detector separations are computed by the proposed method. The channel with shortest source-detector separation and least optical density (*OD*) value does not probe the tissue and is considered as the reference channel. This normalization technique also helps to eliminate common system noise present on all the channels and act as a pre-processing step. Thus, $\mu_a$ values are calculated between the reference channel and all other photon channels for the particular optode geometry using the Eq. (8). Another common problem of cross talk between the $\mu_a$ and $\mu_s'$ is addressed here by accounting for the $L^i$ traveled by detected photons inside a focal region instead of $<L>$ traveled by detected photons as discussed above. Thus, there is no need for any additional baseline measurement to obtain the absolute value for this differential measurement technique.

*2.2 Volumetric 3D Image Reconstruction*

The estimation of photon paths as discussed above and with the assumption that curvature of average photon path is continuous can now act as the basis for image reconstruction. The computed $\mu_a$ values are now back-projected along each of these banana-shaped channels, unlike fan-beam projection where intensity values are simmered in straight lines for image reconstruction. A few points including the minimum point can be selected randomly for each channel and the calculated $\mu_a$ value for each channel is assigned to these pixel locations. The source is rotated around the boundary of the phantom to complete the scanning process, thereby resulting in 208 and 196 photon channels along which the respective calculated μa values are assigned to form the 2-D reconstructed $\mu_a$ image. However, the number of photon path is limited by the source-detector arrangement used for measurement. Thus, the resolution of the reconstructed images is poor. The resolution can be improved by increasing the number of source-detectors but is limited by the dimension of the phantom. Moreover, the data acquisition time will increase considerably with the increase in number of sources and detectors. As such the resolutions of the reconstructed images are improved by interpolation. Here, cubic interpolation is used that interpolates 31 new values in both the x and y-direction between each pair of values in the raw image matrix to form new matrix of dimensions (353×481) and (385×385) for the two phantom geometries considered in this work. Thus the resolution of the reconstructed images can be improved considerably. There is a gradual change in optical properties of biological tissue when looked at mm scale. In literature single Gaussian distribution is used to represent individual heterogeneity [27]. Also, for multiple inclusions or heterogeneity, 2-D multivariate Gaussian Mixture Model (GMM) is proposed to parameterize the optical properties of tissue [28]. So, for reconstruction prior information regarding the presence of inclusions is essential. However, in this method, no such assumption is required.

The reconstructed 2-D profile corresponds to a set of measurement taken at a specific depth z of the phantom. Thus, for the 3-D volumetric imaging, the experiment is repeated a few times at different z values covering the entire depth. A large number of measurement sets can be taken but will considerably increase the data acquisition time. So, here we have made three measurement sets covering the entire depth of the phantoms. Each of these measurements is used by the proposed method for reconstruction of 2-D profiles corresponding to different depths. Now for the 3-D tomographic reconstruction, these 2-D images are stacked together and interpolated over the entire depth of the phantom to reconstruct the 3-D volumetric image [17]. The 3-D image reconstruction is essential for better usability in practical applications and completeness of tomographic reconstruction in its true sense. The volumetric imaging will not only locate the heterogeneity but also helps in observing the penetration depth of the heterogeneity in the phantom volume. Thus, is beneficial for practical disease diagnosis. The imaging across the entire depth of the phantom also gives an idea of the 3-D propagation of photons inside the medium. Photon propagation in tissue like medium is inherently 3-D. So for volumetric imaging 3-D banana function could have been considered by taking simultaneous measurements at different depths of the phantom for 3-D image reconstruction. But that would have increased considerably the computational overhead and the data acquisition time as more number of detector fibers will be needed for data collection. Also, the photon profiles in a highly scattering medium are almost same in all the three planes. A simulation study carried out using the COMSOL Multiphysics shows that the photon propagation is same in all the three planes.

*2.3 Simulation of 3-D Photon propagation*

The results of the 3-D Photon propagation simulation study showed that the photon propagation profiles in all the three planes are same for a simulated wax phantom with CW isotropic light source. COMSOL Multiphysics is used to model the photon propagation in a highly scattering medium. A rectangular slab geometry of dimension (4.4×5×2) cm$^3$ is prepared with material properties similar to wax as discussed below in Sec. 3.1 and is illuminated by a point source at the center of the x-plane. FEM (finite element method) based meshing is done here to divide the entire slab volume into nodes to solve the partial differential equations. In COMSOL for CW domain, the light propagation is modeled by considering the DE in the form of Helmholtz Equation as shown below,

$$\nabla(-c\nabla u) + Au = f \tag{9}$$

where $u = \Phi$ = fluence rate, c = diffusion coefficient, $A$ = 2.74 and $f = S$ = point source.

The 3-D photon propagation profiles are created along the x, y, z planes (considering 10 planes along each direction) and also along the three planes simultaneously as shown in Fig. 1. It is seen that the photon propagation profiles are same in all the three planes. Thus, 3-D images are reconstructed by stacking the 2-D maps reconstructed at different depths and then interpolating it throughout the entire depth of the phantom to complete the volumetric imaging [17].

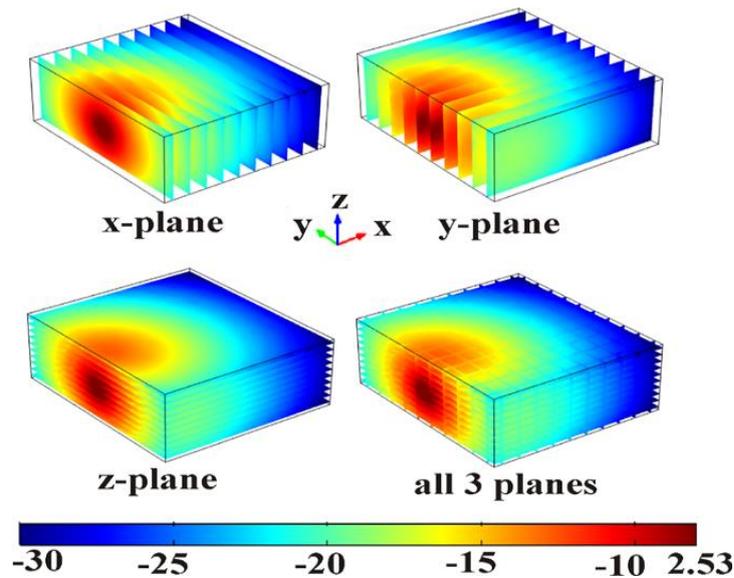

Fig.1. The simulated wax slab geometry with a point source and showing the photon propagation profiles along the three different planes respectively and along all the planes.

*May 21, 2019*

The implementation steps of this semi-analytic method are as follows

Step 1: Data acquisition:
The source fiber sequentially illuminates the phantom at different locations and the attenuated light exiting from the phantom boundary are collected at respective locations by plastic fibers and same is measured with a photo detector.

Step 2: Fitting of non-linear function:
   a. Rosenbrock's function is fitted to all the source-detector channels that estimate the average photon paths followed by the majority photons as curved one.
   a. Estimate the $L^j$ for all the channels.
   b. Apply the MBLL in its proposed form to all the curved photon paths.
   c. Select a reference channel based on OD value and source-detector separation. Calculate the relative $\mu_a$ values between the reference and all other photon paths by using the proposed MBLL in differential form.

Step 3: Semi-analytic tomographic reconstruction from projection
   a. The computed $\mu_a$ values are back-projected along all the curved photon channels for reconstruction of 2-D slice.
   b. The resolution of the 2-D image is improved by interpolation (cubic).
   c. Two more measurements are taken at different depths to have in total three 2-D slices covering the entire depth of the phantom.
   d. All the slices are stacked together one above the other followed by interpolation to reconstruct the final 3-D volumetric image.

*2.4 CS based DOT Reconstruction using Greedy Algorithms*

The imaging physics used in DOT makes detection convenient by utilizing detector arrays like photo detectors. The concept of CS applies significantly less number of measurements for a signal to be recovered from the data [13]. In this scenario, the DOT problem has been converted to single measurement vector (SMV) and multiple measurement vector (MMV) problems using the CS framework [29]. The formulation of DOT forward and inverse problems using SMV and MMV have been discussed in detail using CS framework in our paper [13]. The DOT reconstruction methods with different Greedy algorithms like OMP, CoSaMP, StOMP, ROMP, and S-OMP are studied [29]. Here we considered the CoSaMP algorithms based on SMV formulation for a comparative study [30]. In the CoSaMP algorithm, the 2k sparsity level is used. The algorithm identifies the most significant component of the proxy vectors, and is added to the elements of the current approximation. Finally, it solves the least square step and forms a new approximation vector and update the residual. The main advantage is the stability even with the noisy data. Moreover, it also offers performance guarantee to recover the sparse signal.

## 3. Experimental Setup

The overall experimental setup consists of CW-semiconductor laser, coupled with a multi-mode plastic fiber (diameter 1.5 mm), silicon detector and a customized stage to hold the phantoms. The light from the laser source is guided to the phantom and light exiting from phantom is guided to the detector by 1 mm core diameter multi-mode fibers. The source fiber is rotated around the circumference of the sample along the x–y plane to complete the scanning. This work uses two different phantom geometries (rectangular and cylindrical). For the rectangular phantom experiment, the source fiber is rotated across the boundary at 12 different locations to complete the scanning whereas for cylindrical geometry it is moved in steps at 13 different locations covering the entire periphery of the phantom. The plastic fibers are inserted through holes as shown in Fig. 2. A circular disk holds other ends of detector fibers according to their relative positions to make it convenient for taking boundary measurements. The source-detector fibers are placed along the circumference of imaging plane as shown in Fig. 2. The CW data are collected by using (12 × 16) and (13 × 13) optode arrangements for the two different wax phantom geometries. In both, the cases 16 and 13 numbers of detector fibers are arranged along the boundary of the phantom with a single source fiber to complete a measurement set respectively. Thus, there are in total 208 and 169 measurements respectively. A total of three sets of measurements are taken, and the average is used as final boundary measurement. The experiment is performed in a dark room environment to have background count almost zero. However, the measurements at detector fibers are affected by factors like the proper coupling of the light source to source fiber, the coupling of light from source fiber to detector fiber via sample, and the coupling efficiency of fibers. The proper contact between the sample and fibers are ensured. If there is air-gap between the sample and fiber tip, an anti-reflection coating can be used to reduce energy loss.

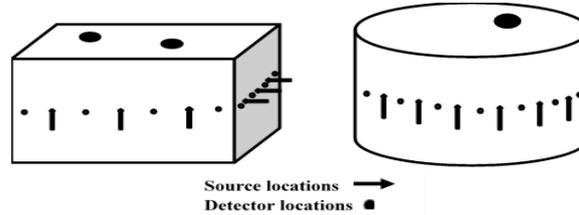

Figure 2: The rectangular and cylindrical wax phantom geometries with inclusions having (12×16) and (13×13) optode arrangement.

Even with necessary precautions, these factors may vary a little while taking measurements. Hence, calibration of data is essential and is explained in our paper [13]. Once calibration is done, the whole experiment is carried out without disturbing the imaging setup. The intensity values at the boundary of the phantom sample are measured by a highly sensitive silicon photodetector with large area head (10×10 mm) and adjustable gain. The lowest noise and highest linearity are achieved by operating the detectors in photovoltaic (unbiased) mode. The experimental setup used here for reconstruction is shown in Fig. 3(a). The setup consists of a CW-semiconductor laser (680 nm), coupled with a multimode plastic fiber (diameter 1.5 mm), silicon detector, circular disk to hold the detector fiber ends and a customized stage to keep the phantom.

*3.1 Preparation of Wax Phantoms*

To mimic the highly diffusive nature of biological tissue cylindrical and rectangular geometry wax phantoms are prepared having background values of $\mu_a$ = 0.25 cm$^{-1}$ and $\mu_s'$ = 20 cm$^{-1}$. The cylindrical phantom has a diameter of 4.75 cm with single inclusion target of diameter 0.70 cm. The rectangular phantom has a dimension of (4.4 × 4.4 × 2.2) cm$^3$ with double anomalous inclusions mimicking the targets (tumors) of diameters 0.40 cm and 0.50 cm respectively. In the homogeneous background inclusions are created by drilling holes up-to depth of z = 2.0 cm and z = 1.5 cm for cylindrical and rectangular phantoms respectively and are filled with various optical contrasts. To mimic optically high absorptive inclusions two different blue Indian ink and glycerin mixtures of 1:1 [v/v] are used with average absorption coefficient values of 5.20 cm$^{-1}$ (78% concentration) and 6.76 cm$^{-1}$ (100% concentration) respectively [28]. The cylindrical phantom uses ink mixture with $\mu_a$ = 5.20 cm$^{-1}$ whereas rectangular uses $\mu_a$ = 6.76 cm$^{-1}$. A homogeneous rectangular wax phantom is also used in this work for reconstruction. Another rectangular wax slab of dimension (4.4 × 5 × 2) cm³ prepared with inclusion at the center covering the entire length of the phantom. The Fig. 3 (b) above shows all the different wax phantoms geometries used in this work.

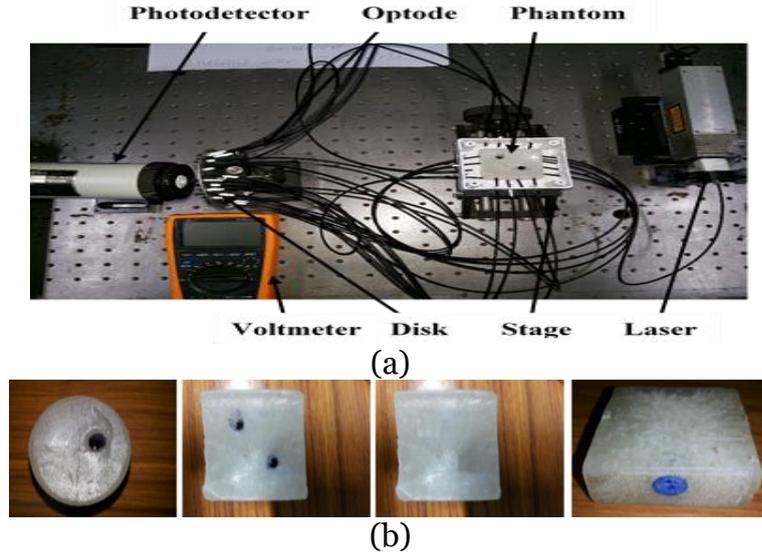

(a)

(b)

Figure 3: The different wax phantoms used are (a) The DOT experimental setup (b) cylindrical, rectangular wax phantoms with single and double inclusion, rectangular homogeneous wax phantom, and rectangular slab wax phantom with inclusion covering the entire length.

## 4. Results and Discussions

*4.1 Experiments on Numerical Phantoms*

To see the effect of geometry on the reconstruction, rectangular and cylindrical numerical phantoms are simulated with optical properties similar to wax phantom experiments. The simulation is performed using TOAST++ package. The solid rectangular phantom has a dimension of (3 × 3) cm with a cylindrical inclusion of diameter 0.5 cm at the center having μ$_a$ value of 6.76 cm$^{-1}$. Whereas, the simulated cylindrical phantom has a diameter of 6 cm with two inclusions mimicking targets with $\mu_a$ values of 5.20 cm$^{-1}$ and 2.72 cm$^{-1}$ respectively. The Fig. 4 (a), (c) shows the simulated phantoms. The simulated data generated by TOAST++ package is used by this method for reconstruction of 2-D absorption profile. The reconstructed absorption images are shown in Fig. 4 (b), (d). It is evident from the simulated and reconstructed images in Fig. 4 (a), (b) and (c), (d) that the proposed method has located absorber accurately. The value of reconstructed $\mu_a$ is 6.81 cm$^{-1}$ for the rectangular phantom and 5.05 cm$^{-1}$ and 2.93 cm$^{-1}$ for cylindrical phantom respectively. These reconstructed values are comparable to the simulated phantoms inclusion values of 6.76 cm$^{-1}$ for rectangular and 5.20 cm$^{-1}$, 2.72 cm$^{-1}$ for the two inclusions (targets) of the cylindrical phantom respectively. However, it is seen as the size of phantom increases it affects the accuracy of reconstructed images. The effect on accuracy can be augmented by increasing number of source-detector pairs.

*May 21, 2019*

The accuracy of reconstructed images is measured by performance evaluation matrix as discussed below in Sec. 4.5.

*4.2 Experiments on Wax Phantoms*

Similarly, to see the effect of geometry in reconstruction two different geometry phantoms as discussed in Sec. 3.1 are considered. The cylindrical phantom has an inclusion center located at (3.20, 2.50) cm of diameter 0.70 cm, whereas the rectangular phantom has a pair of symmetrically positioned inclusions about the center with inclusion centers located at (2.47, 1.70) cm and (1.10, 3.30) cm respectively. The holes are drilled up to depths of 2.0 cm and 1.5 cm respectively. The distance of separations between inclusions is 2.4 cm. The Fig. 5 (a) and (b) shows the 2-D reconstructed images of cylindrical and rectangular wax phantoms at depths of Z = 2.0 cm and Z = 1.0 cm respectively. Here we also imaged the homogeneous (i.e., without any inclusion) rectangular wax phantom geometry to see the reconstruction ability of the proposed method and correctly estimate the absorption change alongside the heterogeneous phantom imaging. The reconstructed 2D image of the homogeneous phantom is shown in Fig. 5 (c). The estimation of homogeneous phantom also gives an idea of the presence of noise in the measurements even after data calibration is done. The calculated $\mu_a$ value for the homogeneous wax phantom is 0.28 cm$^{-1}$ and is comparable to actual $\mu_a$ value of 0.25 cm$^{-1}$ for wax at 680 nm. Thus there is approximately 12% deviation of the reconstructed $\mu_a$ value from the actual one. The Figure 5 (d) shows the comparison of actual quantitative $\mu_a$ values and the reconstructed $\mu_a$ values for the images in 5 (a) and (b) along the transects at y = 2.55 cm; y= 3 cm and y = 1.1 cm respectively. It is clear that the absorption value of the reconstructed image is almost same as that of the actual values (ground truth) for both the phantoms. The bold line represents the reconstructed, and a normal line represents the actual $\mu_a$ values respectively.

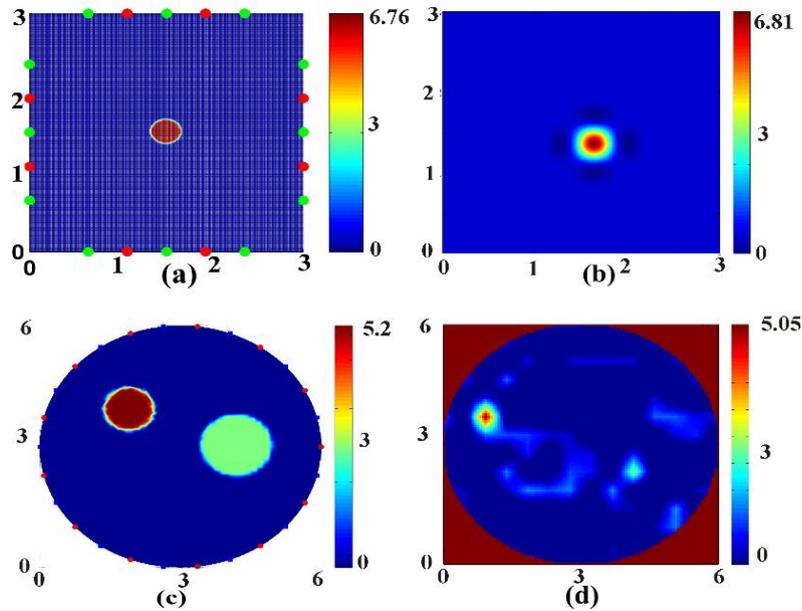

Fig.4. TOAST++ simulated numerical phantoms with single and double inclusions of (a) rectangular, (c) cylindrical geometry phantoms, and the reconstructed 2D absorption profiles of the numerical phantoms with different contrast for the (b) rectangular, and (d) cylindrical geometry.

*May 21, 2019*

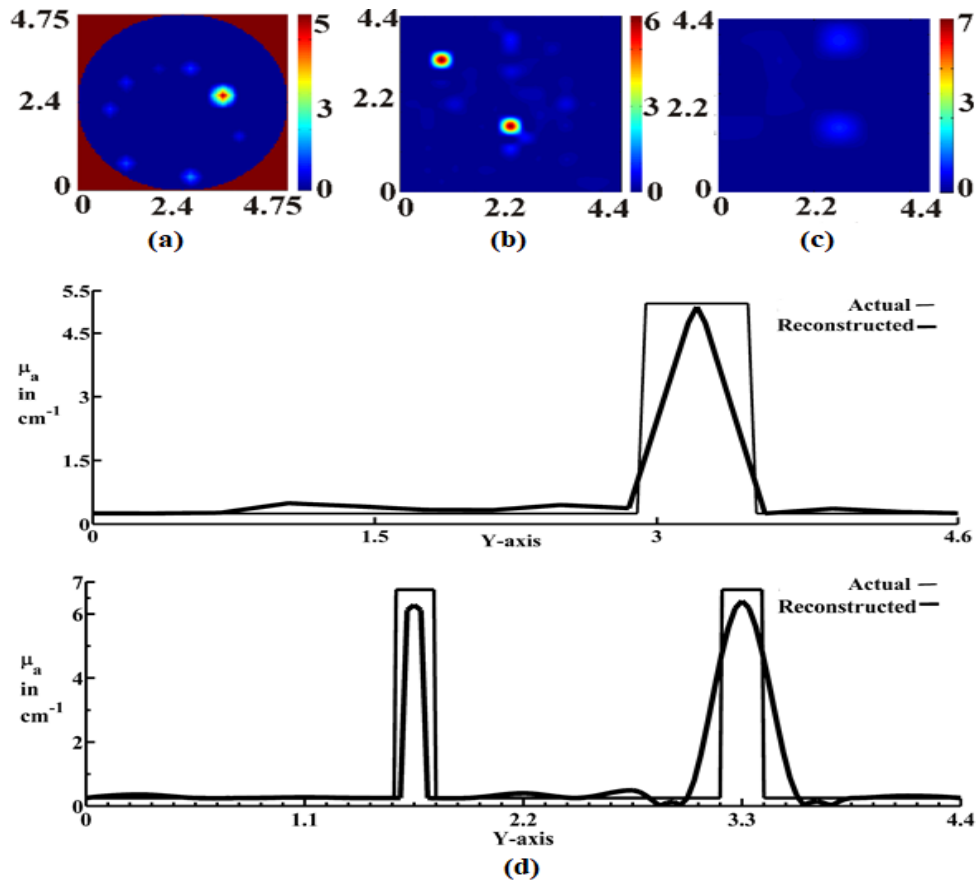

Fig.5. Reconstructed 2-D images of (a) cylindrical single inclusion (b) rectangular two inclusion and (c) homogeneous wax phantoms, and (d) the comparison of actual quantitative absorption coefficient values and reconstructed values for the images in 4 (a) and (b) along the transects at y = 2.55 cm; y= 3 cm, 1.1 cm.

Another slab phantom is used here with the inclusion at the center (blue color portion) as shown in Fig. 6 (a). The source S is placed at a distance of 1 cm towards right from the inclusion. The two detectors $D_1$ and $D_2$ are positioned at equal distances of 1 cm on either side of S, such that $D_1$ falls at the center of the highly absorptive inclusion as shown in Fig. 6 (b) and (c) (image captured on a digital camera). The Fig. 6 (c) is showing the illumination of the phantom by the source S. This experiment is carried out to see the amount of light attenuation that has occurred in the banana-shaped $SD_1$ photon path compared to $SD_2$ path due to the inclusion. As expected there is a 62.04 % increase in the OD values in the first path compared to the second one. Thus, there is a considerable reduction of light in $SD_1$ path compared to the $SD_2$ path due to the presence of absorptive inclusion. This is also observed in the captured image of Fig. 6(c).

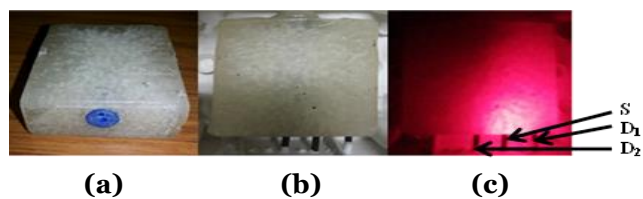

Fig.6. The images (captured in a digital camera) of the (a) slab phantom with the inclusion at center in blue color, (b) showing the source (S) and two detectors ($D_1$, $D_2$) on either side of the S, and (c) the same setup with the source (S) illuminating the phantom.

*4.3 Volumetric 3-D Reconstruction*

For better usability and completeness, this method is extended to 3-D volumetric imaging for both the rectangular and cylindrical wax phantoms. For 3-D imaging three sets of measurements are taken at different depths of Z = 0.5 cm, Z = 1.1 cm, Z = 1.8 cm for rectangular and Z = 1.0 cm, Z = 2.0 cm, Z = 3.0 cm for the cylindrical phantoms respectively. The Fig. 7(a)-(c) and (d)-(f) shows the 2-D absorption profiles at these different depths for the rectangular and cylindrical phantoms respectively. The images in Fig. 7(a)-(c) are the raw non-interpolated images whereas (d)-(f) are the interpolated images. The raw images clearly showing the different values being spread out across the photon paths. The interpolation showed considerable improvement in the resolution of the reconstructed images as discussed in Sec. 2.2. The Fig. 7(g) and (h) shows the reconstructed 3-D volumetric images of both the rectangular and cylindrical phantoms respectively. The holes are drilled to a depth of 1.5 cm and 2.0 cm from the top for both the phantom geometries. Interestingly, from the 2-D reconstructed image at Z = 0.5 cm and Z = 1.0 cm from the bottom for both the phantom geometries, the absorption profiles are almost homogeneous and with lower absorption values. The inclusions are drilled up to a depth of 1.5 cm and 2.0 cm from the top only as such no inclusion is present at a depth of 0.5 cm and 1.0 cm. The reconstructed images reflected this, and there is no effect of inclusion as such. Whereas, in the other two reconstructed images at depths of 1.1 cm, 1.8 cm, and 2.0 cm, 3.0 cm, the targets (red portion) are easily visible showing the presence of inclusions. Thus, this simple and computationally efficient imaging technique has good recovery accuracy for real-time DOT applications that demand fast reconstruction.

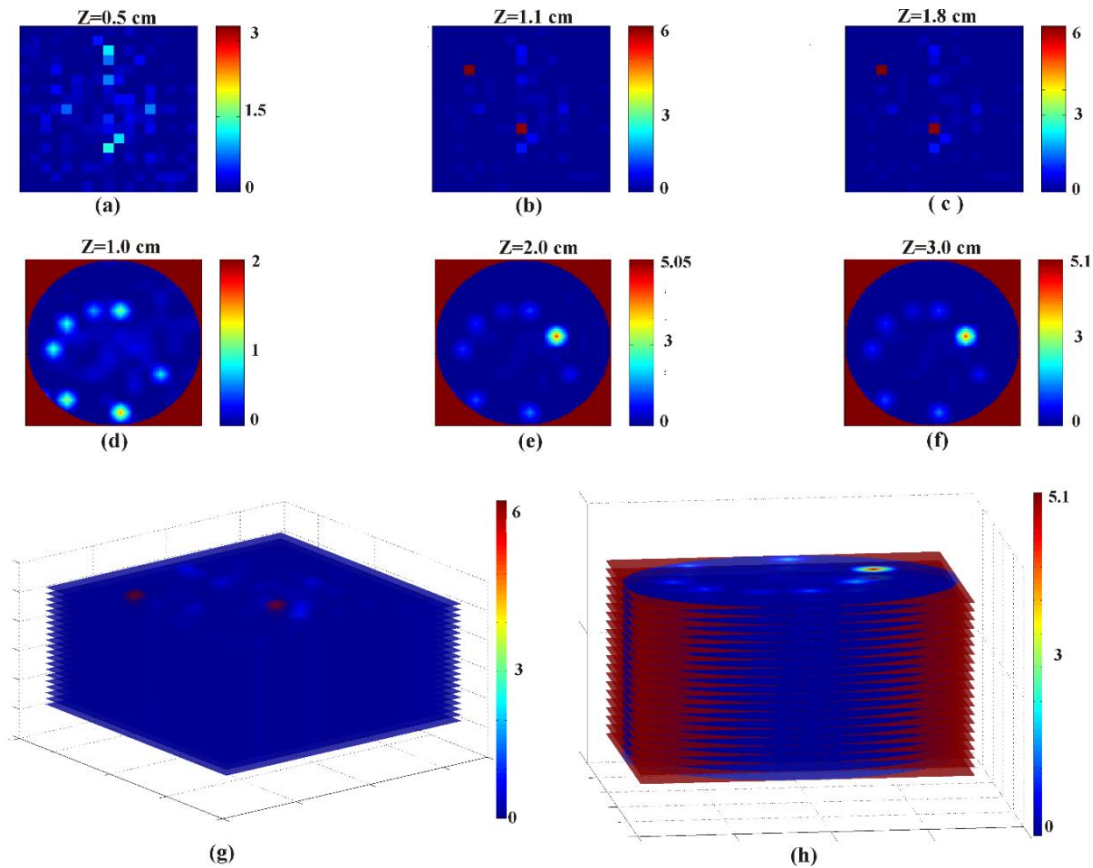

Fig.7. Reconstructed 2-D images (a) - (c) of two inclusions rectangular wax phantom at three different depths, (d) - (f) of one inclusion cylindrical wax phantom at three different depths, reconstructed 3-D volumetric $\mu_a$ images of (g) the rectangular and (h) the cylindrical wax phantoms respectively.

*4.4 Comparison with CoSAMP, and SA-MUSIC algorithm*

This method is compared with the CoSAMP as explained in sec.2.3 and the SA-MUSIC algorithm. The comparative study is carried out on rectangular wax phantoms with both single and double inclusion respectively. The Figure 8(a) and (b) below shows the reconstructed 2-D images of rectangular wax phantoms with a single and double inclusion by this method. Whereas, the Fig. 8(c) and (d) are the reconstructed images of the same phantoms with the CoSAMP, and the SA-MUSIC algorithms respectively. The comparison of reconstructed images shows that the contrast is almost similar, but the curved beam method is more accurate in locating targets compared to the CoSaMP. The CoSaMP algorithm gives better contrast but fails to estimate the depth of the inclusion correctly. So for the double inclusion case, the SA-MUSIC algorithm is used that located the inclusions rather accurately. However, the curved beam method is even more accurate than the SA-MUSIC algorithm. Also, the reconstructed $\mu_a$ values for this method are closer to the ground truth (ink $\mu_a$) values compared to both the CoSaMP and the SA-MUSIC algorithms. The comparison shows the potential of this method in DOT with acceptable accuracy using a much simpler, efficient, and effective approach.

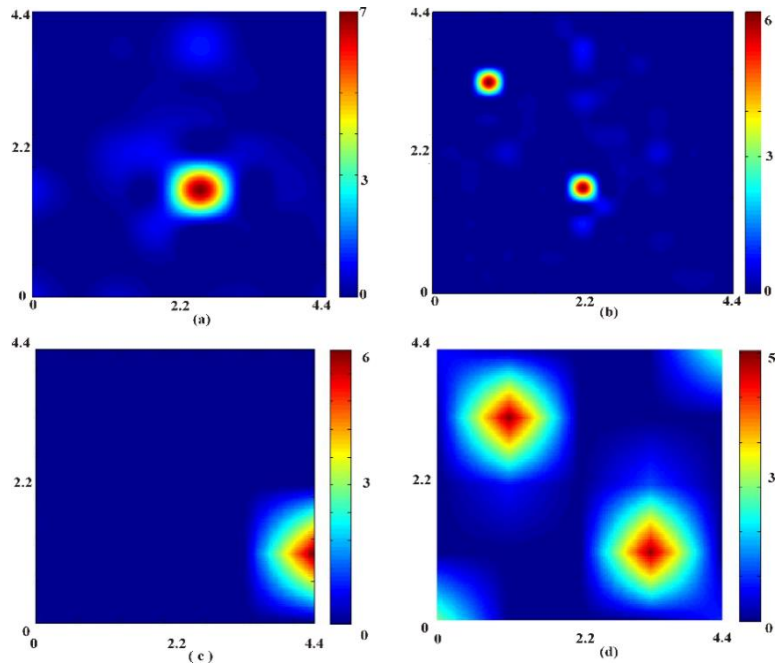

Fig.8. Comparison of reconstructed 2-D images by this method of the same rectangular wax phantoms (a), (c) single inclusion with the CoSaMP algorithm and (b), (d) two inclusions with the SA-MUSIC algorithm respectively.

*4.5 Accuracy Estimation*

The accuracy in locating inclusions can be estimated by comparing the inclusion locations in the reconstructed images with the actual inclusion locations in the simulated and wax phantoms used in this work. Table 1 shows the inclusion center of TOAST++ simulated and the reconstructed images of numerical phantoms by this method. Table 2

*May 21, 2019*

also gives the comparison of inclusion centers in the reconstructed images and the actual wax phantoms (ground truths) of different geometry. Table 3 is showing a comparison of reconstruction accuracy of this curved-beam method with that of the CoSaMP and SA-MUSIC algorithms for inclusion location and $\mu_a$ values in the reconstructed images. These tables show the correlation of the reconstructed and actual $\mu_a$ value of ink used in the inclusion of the wax phantom as well as for TOAST++ simulated numerical phantom experiments. It is seen from both these tables that this method has satisfactorily reconstructed $\mu_a$ values and located the inclusion centers for both the numerical and wax phantom experiments.

**Table 1.** Comparison of TOAST++ simulated and reconstructed images of numerical phantoms.

| Parameters | Simulated rectangular | Reconstructed rectangular | Simulated cylindrical | Reconstructed cylindrical |
|---|---|---|---|---|
| Location (x, y) in cm | (1.5, 1.5) | (1.6, 1.5) | (1.5, 4.0) and (4.5, 3.0) | (1.15, 3.85) and (4.40, 2.75) |
| $\mu_a$ value in cm$^{-1}$ | 6.76 | 6.81 | 5.20 and 2.72 | 5.05 and 2.93 |

**Table 2.** Comparison of ground truth and reconstructed images of wax phantoms.

| Parameters | Ground truth rectangular | Reconstructed rectangular | Ground truth cylindrical | Reconstructed cylindrical |
|---|---|---|---|---|
| Location (x, y) in cm | (2.47, 1.70) and (1.10, 3.30) | (2.35, 1.70) and (0.93, 3.30) | (3.20, 2.50) | (3.25, 2.60) |
| $\mu_a$ value in cm$^{-1}$ | 6.76 | 6.28 and 6.40 | 5.20 | 5.10 |

**Table 3.** Comparison of reconstructed images by the semi-analytic with the CoSaMP and Sa-MUSIC.

| Parameters | Semi-anallytic method | | CoSaMP | SA-MUSIC |
|---|---|---|---|---|
| | Rectangular (1 inclusion) | Rectangular (2 inclusion) | Rectangular (1 inclusion) | Rectangular (2 inclusion) |
| Location (x, y) cm | (2.35, 1.70) | (2.35, 1.70) & (0.95, 3.30) | (4.25, 1.20) | (4.25, 1.20) & (2.15, 4.25) |
| $\mu_a$ value cm$^{-1}$ | 6.83 | 6.28 & 6.40 | 6.40 | 5.15 & 5.45 |

The performance evaluation metrics play an important role in accessing performance of DOT reconstruction methods. Here different performance metrics like structural similarity index (SSIM), mean squared error (MSE), and peak signal to noise ratio (PSNR) are evaluated to assess the performance of the reconstructed images by this method. The SSIM is calculated to show the structural similarity of reconstructed images and the actual images. For any two images, the SSIM is computed as follows [20]

$$S(x,y) = \frac{(2\mu_x\mu_y+C_1)(2\sigma_{xy}+C_2)}{(\mu_x^2+\mu_y^2+C_2)(\sigma_x^2+\sigma_y^2+C_2)} \qquad (10)$$

where $\mu_x$, $\mu_y$, $\sigma_{xy}$, $\sigma_x$, $\sigma_y$ are the mean, covariance and standard deviation of the two images x and y. The decimal value is between –1 and +1. The higher is the positive value of SSIM more is the structural similarity. Also, the MSE and PSNR are calculated to assess the performance of the reconstructed images as follows

$$\frac{1}{N}\sum_{i=1}^{N}(\mu_a^r - \mu_a^t)^2 \qquad (11)$$

where $\mu_a^r$ and $\mu_a^t$ are the reconstructed and true absorption coefficient values, $N$ is the number of absorption coefficient values.

$$10\log_{10}\left(\frac{(peakval)^2}{MSE}\right) \qquad (12)$$

where *peakval* is the peak value taken from the range of the image.

The MSE and PSNR are the two error metrics used to compare image quality. The MSE represents the cumulative squared error between the reconstructed and the original image, whereas PSNR represents a measure of the peak error and is measured in decibels (dB). The lower the value of MSE, lower is the error and higher the PSNR, the better is the quality of the reconstructed image.

The different performance metrics calculated for the absorption images reconstructed by this non-linear method for numerical and wax phantoms are shown below in Table 4 for two different phantom geometries considered here in this paper. Irrespective of the geometry, the overall comparisons of performance evaluation metrics for the reconstructed images for both the numerical and wax phantoms experiments are satisfactory. Also, the values of MSE, PSNR, and SSIM of the reconstructed images for this non-linear method have outperformed the CoSaMP and the state-of-art SA-MUSIC algorithms for the rectangular phantoms with single and double inclusions. The MSE values are less showing good contrast whereas the SSIM and PSNR of the reconstructed images are high showing good correlation between the reconstructed and the ground truth images as shown in Table 4. A high value of SSIM signifies less structural deformity, i.e., the artifacts are less in the reconstructed images.

**Table 4.** A comparison of the performance evaluation metrics of reconstructed images by this semi-analytic method with the CoSaMP and SA-MUSIC algorithms.

| Performance Metrics | Reconstructed image (numerical phantom) | | Reconstructed image (wax phantom) | | Reconstructed image (wax phantom) | |
|---|---|---|---|---|---|---|
| | Curved-beam | | Curved-beam | | CoSaMP | SA-MUSIC |
| | Rectangular 1 inclusion | Cylindrical 2 inclusion | Rectangular 2 inclusion | Cylindrical 1 inclusion | Rectangular 1 inclusion | Rectangular 2 inclusion |
| MSE | 0.033 | 0.052 | 0.049 | 0.034 | 0.057 | 0.063 |
| SSIM | 0.792 | 0.573 | 0.649 | 0.732 | 0.520 | 0.591 |
| PSNR(dB) | 33.203 | 27.102 | 29.201 | 31.701 | 25.311 | 27.201 |

## 5. Conclusion

The non-recursive linear algorithms have been used in diffuse optical imaging with simple linear relationship to estimate the photon paths [14]. The linear method does not solve the inverse problem, but due to the linear relationship, there is distortion in the reconstructed images. This paper presents a CW non-recursive and non-linear method for 3-D reconstruction by fitting a semi-analytic Rosenbrock's banana function to approximate the photon path for source-detector channels. The tissue volumes interrogated by these overlapping banana-shaped channels are sensitive to absorption change at different depths. The potential of this non-linear method in DOT reconstruction is studied by considering extensive numerical simulations and wax phantom experiments with different geometries and inclusions. The calculated $\mu_a$ values are back projected along the curved channels and serve as imaging operator for the 3-D tomographic reconstruction. Thus there is no need to solve the inverse problem like the non-recursive linear algorithm. However, fitting of the non-linear function estimates the photon paths as curved one and thus reduces image distortion present in the linear reconstruction method. Another advantage of this technique is that there is no need to assume the inclusion location, shape and size.

The 3-D reconstructed images of numerical simulations and wax phantom experiments with different geometry and contrast show the potential and validate this non-linear method for DOT application. This 3-D tomographic imaging method completes the reconstruction in its true sense as well as gives better usability in real-time DOT applications. The accuracy of reconstructed images is evaluated by using MSE, SSIM, and PSNR. The comparative study revealed that the recovery accuracy is better than the mathematically exhaustive Greedy algorithms. Hence this proposed method is stable, accurate, and computationally efficient when compared to state-of-art DOT reconstruction methods. Also, there is scope to use the images reconstructed by this method as seed image for the recursive CS based methods to improve accuracy and efficiency. The amalgamation of both these methods needs to be explored and may lead to development of improved and more efficient DOT reconstruction algorithms for real-time DOT applications. The future work may be carried out in this direction.

*May 21, 2019*